# A Disintegrating Minor Planet Transiting a White Dwarf

Andrew Vanderburg[1], John Asher Johnson[1], Saul Rappaport[2], Allyson Bieryla[1], Jonathan Irwin[1], John Arban Lewis[1], David Kipping[1,3], Warren R. Brown[1], Patrick Dufour[4], David R. Ciardi[5], Ruth Angus[1,6], Laura Schaefer[1], David W. Latham[1], David Charbonneau[1], Charles Beichman[5], Jason Eastman[1], Nate McCrady[7], Robert A. Wittenmyer[8], & Jason T. Wright[9,10].

**White dwarfs are the end state of most stars, including the Sun, after they exhaust their nuclear fuel. Between 1/4 and 1/2 of white dwarfs have elements heavier than helium in their atmospheres[1,2], even though these elements should rapidly settle into the stellar interiors unless they are occasionally replenished[3–5]. The abundance ratios of heavy elements in white dwarf atmospheres are similar to rocky bodies in the Solar system[6,7]. This and the existence of warm dusty debris disks[8–13] around about 4% of white dwarfs[14–16] suggest that rocky debris from white dwarf progenitors' planetary systems occasionally pollute the stars' atmospheres[17]. The total accreted mass can be comparable to that of large asteroids in the solar system[1]. However, the process of disrupting planetary material has not yet been observed. Here, we report observations of a white dwarf being transited by at least one and likely multiple disintegrating planetesimals with periods ranging from 4.5 hours to 4.9 hours. The strongest transit signals occur every 4.5 hours and exhibit varying depths up to 40% and asymmetric profiles, indicative of a small object with a cometary tail of dusty effluent material. The star hosts a dusty debris disk and the star's spectrum shows prominent lines from heavy elements like magnesium, aluminium, silicon, calcium, iron, and nickel. This system provides evidence that heavy element pollution of white dwarfs can originate from disrupted rocky bodies such as asteroids and minor planets.**

WD 1145+017 (also designated EPIC 201563164) is a helium-envelope white dwarf (Table S1) that was observed by NASA's *Kepler* Space Telescope during the first campaign of its two-wheeled mission, hereafter referred to as K2. After processing K2 data for WD 1145+017 to produce a light curve and correct for instrumental systematics[18], we identified a transit-like signal with a period of 4.5 hours using a box-fitting least-squares search algorithm[19]. Using a Fourier analysis on the systematic-corrected K2 data, we identified five additional weaker, but statistically significant, periodicities in the data, all with periods between 4.5 and 5 hours (Figure 1, Table S2). We examined the dominant periodicity and found that the depth and shape of the transits varied significantly over the 80 days of K2 observations (Figure 2).

We initiated follow-up ground-based photometry to better time-resolve the transits seen in the K2 data (Figure S1). We observed WD 1145+017 frequently over the course of about a month with the 1.2-meter telescope at the Fred L. Whipple Observatory (FLWO) on Mt. Hopkins, Arizona; one of the 0.7-meter MINERVA telescopes, also at FLWO; and four of the eight 0.4-meter telescopes that compose the MEarth-South Array at Cerro Tololo Inter-American Observatory in Chile. Most of these data showed no interesting or significant signals, but on two nights we observed deep (up to 40%), short-duration (5 minutes), asymmetric transits separated by the dominant 4.5 hour period identified in the K2 data (Figure 3). In particular, we detected two transits 4.5 hours apart on the night of 11 April, 2015 with the 1.2-meter FLWO telescope in V-band (green visible light) and two transits separated by the same 4.5 hour period with four of the eight MEarth-South array telescopes on the night of 17 April, 2015, all in near infrared light (using a 715 nm long pass filter). The transits did not occur at the times predicted from the K2 ephemeris, and the two transits detected on April 11 happened nearly 180 degrees out of phase from the two transits detected on April 17. Observations with MEarth-South in near–infrared light and with MINERVA in white visible light the next night (April 18) showed only a possible 10-15% depth transit event in phase with the previous night's events. The 5-minute duration of the transits is longer than the roughly 1-minute duration we would expect for a solid body transiting the white dwarf. We confirmed that these events are in fact transits of the white dwarf. The depth and morphology of the transits we see in the ground–based data cannot be explained by stellar pulsations, and archival and adaptive optics imaging place strong constraints on blend scenarios involving a background eclipsing binary (Figure S2).

We also obtained spectroscopic observations with the MMT Blue Channel spectrograph which we used to place limits on radial velocity (RV) variations that would indicate stellar companions. The RV measurements exclude companions larger than 10 Jupiter masses at the 95% confidence level.

The spectra also reveal that the envelope of the white dwarf contains magnesium, aluminium, silicon, calcium, iron, and nickel (Figure S3). These elements heavier than helium have settling times much shorter than the cooling age of the white dwarf, indicating that they have been deposited in the white dwarf's envelope in the last million years[5], much more recently than its formation about 175 ± 75 million years ago. Archival photometry for this system is well fitted by a 15,900 K metal-rich white dwarf model spectrum, and we find evidence for excess infrared emission consistent with a warm (1,150 K) dusty debris disk (Figure S4).

[1]Harvard-Smithsonian Center for Astrophysics, Cambridge, MA 02138 USA. [2]Department of Physics, and Kavli Institute for Astrophysics and Space Research, Massachusetts Institute of Technology, Cambridge, MA 02139, USA. [3]Department of Astronomy, Columbia University, New York, NY 10027, USA. [4]Institut de Recherche sur les Exoplanètes (iREx), Départment de Physique, Université de Montréal, Montréal, QC H3C 3J7, Canada. [5]NASA Exoplanet Science Institute, California Institute of Technology, Pasadena, CA 91125, USA. [6]Department of Physics, University of Oxford, Oxford OX1 3RH, UK. [7]Department of Physics and Astronomy, University of Montana, Missoula, MT 59812 USA. [8]School of Physics and Australian Centre for Astrobiology, University of New South Wales, Sydney, NSW 2052, Australia. [9]Department of Astronomy and Astrophysics and Center for Exoplanets and Habitable Worlds, The Pennsylvania State University, University Park, PA 16802. [10]NASA Nexus for Exoplanet System Science

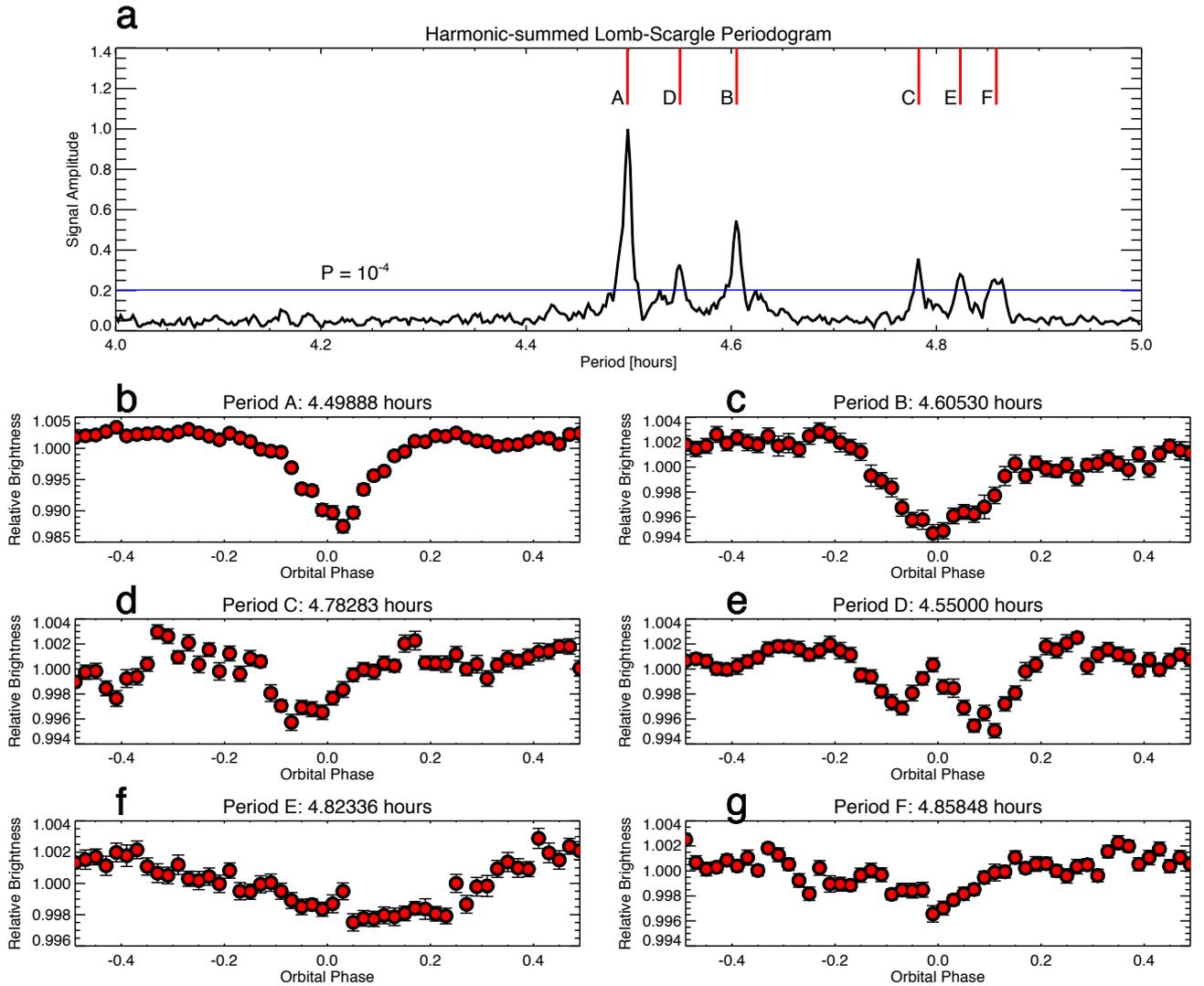

**Figure 1: Six significant periodicities found in the K2 data.** Panel a: Harmonic-summed Lomb-Scargle periodogram of the K2 data. Panels b-g: K2 light curve folded on the six significant peaks and binned in phase. We label the signals A-F in order of significance and plot folded light curves in the lower panels. When plotting each fold, we sequentially removed stronger signals by dividing the dataset by the binned, phase folded light curves of the stronger signals. Note the differences in y scale on the individual panels. Error bars shown are the standard errors of the mean within each bin.

We interpret these observations as evidence for at least one, and likely six or more disintegrating planetesimals transiting a white dwarf. Disintegrating planets have been observed transiting main sequence stars[20–22], and show asymmetric transit profiles and variable transit depths, similar behaviours to those which we see here. These previously detected disintegrating planets are believed to be heated by the host star and losing mass through Parker-type thermal winds, the molecules in which condense into the obscuring dust observed to be occulting the star[23]. The solid bodies themselves are too small to detect, so the transits are dominated by the much larger dust cloud trailing the planets. The density of the dust cloud is in presumed to be highly variable, which gives rise to the variable transit depths, and comet-like dust tails can explain the asymmetric transit shapes[20–22]. In the case of WD 1145+017 we identify six stable periodicities in the K2 light curve which could be explained by occultations of the central star by dust clouds. We propose that each of these periodicities could be related to individual planetesimals (or multiple fragments of one minor planet) orbiting the white dwarf star near the tidal radius for rocky bodies. Each of these planetesimals would sporadically launch winds of metal gases that are most likely freely streaming from the body and which condense into dust clouds that periodically block the light of the white dwarf. A trailing dust cloud explains the variable transit depths, asymmetric transit profiles, and longer-than-expected transit durations we see in the light curves of WD 1145+017 (Figure S5).

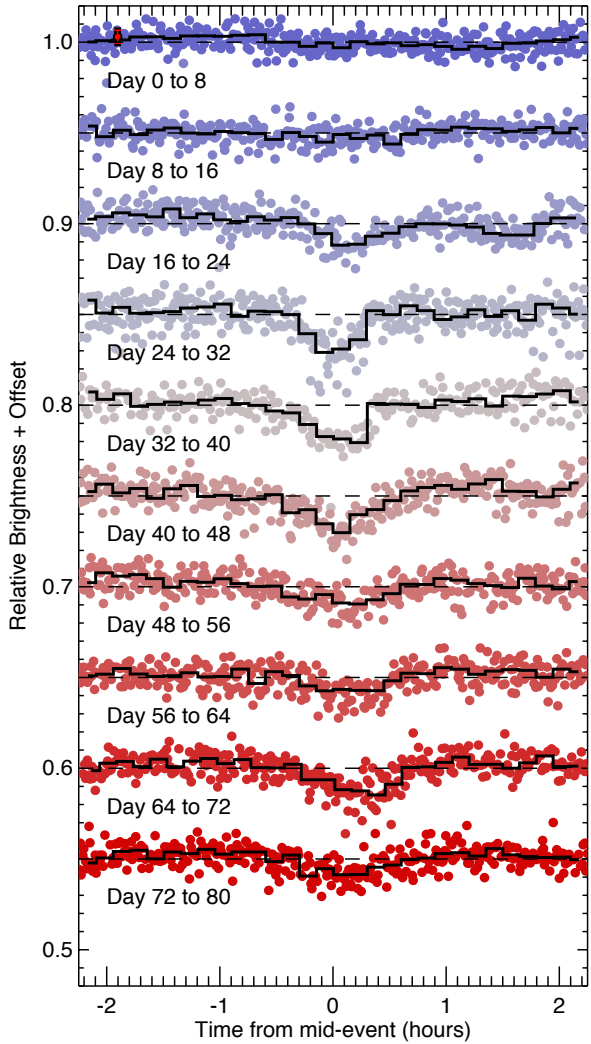

**Figure 2: Evolution of the K2 transit light curve over 80 days of observations.** We show the K2 light curve broken into segments 8 days in length and folded on the most significant, 4.5 hour period. The individual datapoints (sampled with a 30 minute integration time) are shown as dots, with the colour representing the segment in time. The averaged light curve for each bin is shown as a solid black line. Each light curve segment is vertically offset for clarity. We show the typical measurement uncertainty (standard deviation) with a red error bar on one datapoint in the upper left.

We have simulated the dynamics of six planetesimals in circular orbits with periods between 4.5 and 4.9 hours and find that such a configuration is stable for at least $10^6$ orbits provided that their masses are smaller than or comparable to Ceres ($1.6 \times 10^{-4}$ $M_{\oplus}$) or possibly Haumea ($6.7 \times 10^{-4}$ $M_{\oplus}$). These six planetesimals must be rocky (because gaseous bodies would overflow their Roche lobes), and must have densities greater than $\rho \gtrsim 2$ g cm$^{-3}$ to not be tidally disrupted in such short period orbits[24,25]. We also simulated the dynamics of different planetesimals in 1:1 mean motion orbital resonances (for example, horseshoe orbits), and find that two different planetesimals in such orbits outbursting at different times could plausibly explain the difference in orbital phases between the K2 light curve, the April 11 events, and the April 17 events.

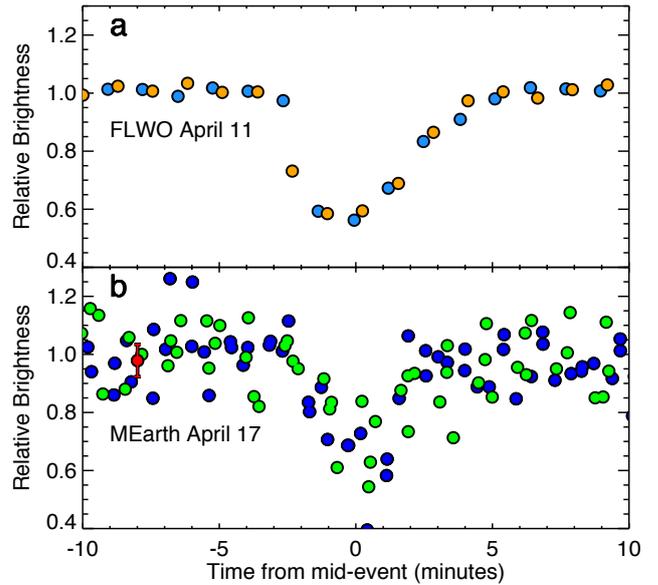

**Figure 3: Transit light curves measured from two ground-based facilities.** Panel a: Two events observed at FLWO with a separation equal to the 4.5 hour A-period detected by K2. The first FLWO event is blue, and the second is orange. Panel b: Two events observed by MEarth-South separated by the 4.5 hour A-period. The first MEarth-S event is blue, and the second is green. The typical MEarth-S measurement uncertainty (standard deviation) is shown as a red error bar on one datapoint. The FLWO error bars are smaller than the size of the symbols.

We estimate that a mass loss rate of roughly $8 \times 10^9$ g s$^{-1}$ is necessary to explain the transits we see. Various refractory materials (including iron, fayalite, albite, and orthoclase) heated by the white dwarf could plausibly sublimate from a planetesimal roughly the size of Ceres at this rate despite the white dwarf's relatively low luminosity (Figure S6). These metal vapours would be quickly lost via free-streaming winds or Jeans escape, since the planetesimal escape velocity is comparable to the metal vapour's thermal speed. We simulated a dust cloud condensed from the escaped metal vapour in orbit[20–22] and find that the radiation environment in which these planetesimals are situated can give rise to dust tails like we infer from the ground-based transit observations (Figure S7). Collisions with disk debris[26] could also plausibly cause mass from the planetesimal to be lost into orbit.

A possible formation scenario is one that involves minor planets that are left over from the progenitor stellar system before the star evolved[4,17]. In this scenario, mass loss from the host star disturbs the stability of the planetary system, which can lead to planets or smaller objects like asteroids or comets being scattered inwards to orbital radii much smaller than the size of the progenitor star when it was an evolving giant. A challenge for this model is placing the planetesimals in close concentric orbits so near the star without being totally disrupted. Current models suggest that planetesimals can be scattered inwards on highly eccentric orbits, tidally disrupted into elliptical dust disks, and circularised by Poynting-Robertson drag[27]. Bodies that can release enough dust to cause the transits of WD 1145+017 are too large to be circularised in this way. Recent theoretical work[28] studying smaller bodies has shown that outgassing material can quickly circularise orbits, but it is unclear how this process scales to the massive bodies inferred here.

We note that our interpretation of this system is still uncertain. In particular, it is hard to explain the phase shifts between the ground-based transits and the K2 transits, and more ground-based observations are necessary to understand this effect. Another possible model for the system is that small rings[27] or debris clouds of disrupted planetary material in a disk occasionally cross in front of the star and block its light. While this could explain the large phase shifts we see between the FLWO and MEarth transits, it is difficult to explain the highly stable periods ($\Delta P/P \lesssim 10^{-4}$) seen in the K2 data without massive orbiting bodies (Figure S8). Fortunately, the large transit depths make follow-up observations that could distinguish among these scenarios feasible both from the ground and space. It might be possible to detect periodic infrared emission[29] from the orbiting planetesimals with the James Webb Space Telescope. Additional follow-up observations such as transit spectroscopy could constrain both scenarios by detecting the presence of molecules in the dust tails or the wavelength dependence of the dust scattering[30].

The observations presented in this paper, in particular the heavy element pollution of the white dwarf, evidence for a warm dusty debris disk, and transits of disintegrating planetesimals, are consistent with a scenario suggested over the past decade in which the orbits of rocky bodies are occasionally perturbed and pass close enough to white dwarf stars to become tidally disrupted, leading to the infall of debris onto the star's surface. Observations have shown that this scenario could be quite common among white dwarfs, with between 25% and 50% of white dwarfs showing evidence for heavy element pollution. This detection indicates that disintegrating planetesimals may be common as well (Figure S9). The transits of WD 1145+017 provide evidence of rocky, disintegrating bodies around a white dwarf and support the planetesimal accretion model of polluted white dwarfs.

**Acknowledgements** We thank Bryce Croll, Dimitri Veras, Matt Holman, Ryan Loomis, Juliette Becker, Kat Deck, Hilke Schlichting, Henry Lin, Avi Loeb, and David Osip for valuable discussions and assistance. We thank Carter Allinson, Sean Dillet, Danielle Frostig, April Johnson, Daniel Hellstrom, Steven Johnson, Barra Peak, and Taylor Reneau for conducting MINERVA observations. We thank Mark Wyatt for suggesting the method of presentation in Figure S8. AV is supported by the National Science Foundation Graduate Research Fellowship, Grant No. DGE 1144152. J.A.J is supported by generous grants from the David and Lucile Packard Foundation and the Alfred P. Sloan Foundation. The Center for Exoplanets and Habitable Worlds is supported by the Pennsylvania State University, the Eberly College of Science, and the Pennsylvania Space Grant Consortium. The MEarth Team gratefully acknowledges funding from the David and Lucile Packard Fellowship for Science and Engineering (awarded to D.C.), the National Science Foundation under grants AST-0807690, AST-1109468, and AST-1004488 (Alan T. Waterman Award), and a grant from the John Templeton Foundation. The opinions expressed in this publication are those of the authors and do not necessarily reflect the views of the John Templeton Foundation. This research has made use of NASA's Astrophysics Data System, the SIMBAD database and VizieR catalog access tool operated at CDS, Strasbourg, France. Some of the data presented in this paper were obtained from the Mikulski Archive for Space Telescopes (MAST). This paper includes data from the *Kepler*/K2 mission, the Wide-field Infrared Survey Explorer, the MMT Observatory, the Sloan Digital Sky Survey (SDSS-III), the National Geographic Society-Palomar Observatory Sky Atlas (POSS-I) and the W.M. Keck Observatory. MINERVA is made possible by generous contributions from its collaborating institutions and Mt. Cuba Astronomical Foundation, the David and Lucile Packard Foundation, the National Aeronautics and Space Administration, and the Australian Research Council. The authors wish to recognise and acknowledge the very significant cultural role and reverence that the summit of Maunakea has always had within the indigenous Hawai'ian community. We are most fortunate to have the opportunity to conduct observations from this mountain.

**Competing Interests** The authors declare that they have no competing financial interests.

**Author contribution** AV processed and searched the K2 data, identified this system, analysed the K2 data for WD 1145+017 (with help from SR, DK, and JTW), processed the MINERVA data, measured radial velocities (with help from WRB and DWL), and was the primary author of the manuscript. SR performed the dynamical calculations and dust simulations. WRB obtained and reduced the MMT spectra. PD analysed the MMT spectra and SDSS photometry to measure spectroscopic properties. JAL analysed photometry and modelled the excess infrared emission. AB and DWL obtained and processed the FLWO data. JI and DC obtained and processed the MEarth data. DRC and CB obtained and processed the Keck data. RA calculated the systematics insensitive periodogram. LS calculated vapour pressures for some minerals with MAGMA. JAJ, JE, NM, RAW, and JTW made using MINERVA possible. JAJ provided scientific leadership.

**Correspondence** Correspondence and requests for materials should be addressed to AV (email: avanderburg@cfa.harvard.edu).

**Reprints** Reprints and permissions information is available at www.nature.com/reprints

**Data Deposition** The raw K2 data is available at http://archive.stsci.edu/k2/data_search/search.php under the K2 identification number 201563164. The processed K2 data is available at https://archive.stsci.edu/missions/hlsp/k2sff/html/c01/ep201563164.html .

# Supplementary Information:
# A Disintegrating Minor Planet Transiting a White Dwarf

Andrew Vanderburg, John Asher Johnson, Saul Rappaport, Allyson Bieryla, Jonathan Irwin, John Arban Lewis, David Kipping, Warren R. Brown, Patrick Dufour, David R. Ciardi, Ruth Angus, Laura Schaefer, David W. Latham, David Charbonneau, Charles Beichman, Jason Eastman, Nate McCrady, Robert A. Wittenmyer, & Jason T. Wright

**K2 Light Curve Preparation** NASA's K2 mission[31] has not produced photometric light curves from most of its data, so we began by downloading pixel–level time series data from the Mikulski Archive for Space Telescopes (MAST). We processed the data in a way similar to that proposed in a previous study[18]. In brief, we performed simple aperture photometry on the K2 pixel time series of WD 1145+017 with a set of 20 different apertures of different shapes and sizes. Raw aperture photometry from K2 is dominated by systematic noise caused by the motion of the spacecraft, which is no longer able to point precisely due to the failure of 2 of its 4 reaction wheels. Instead, *Kepler* uses its two functional reaction wheels to point the centre of its field of view and balances its solar panels opposite the Sun. This configuration is unstable, so *Kepler* rolls slightly about the centre of its field of view on the timescale of hours, after which time the roll is corrected by thruster fires.

To correct for *Kepler*'s motion, after extracting photometry, we decorrelated the roll of the spacecraft (measured by calculating the centroid position of the bright star EPIC 201611708) with the measured flux using an iterative technique. The biggest difference between our procedure and that used in the previous study[18] is the duration of the observations. The previous study was based on only 6.5 days of data from an engineering test of the K2 operating mode and was able to treat the motion of the spacecraft as a one-dimensional arc in the roll direction. Over the 80 days of observations we processed, there was significant drift of the images transverse to the roll of the spacecraft, likely due to differential velocity aberration. To compensate for this transverse motion, we broke up the time series into six different chunks about 13 days in length. Over this shorter time period, the motion transverse to the roll of the spacecraft is negligible, and we were able to decorrelate the photometry from the roll of the spacecraft as done previously[18]. We identified datapoints that were collected while *Kepler*'s thrusters were firing and excluded them, as those data can exhibit anomalous behaviour[18]. After correcting for the motion of the spacecraft in the light curves from each of the 20 different photometric apertures, we calculated each light curve's photometric precision and chose the one yielding the best precision. We tested to make sure that our choice of aperture did not substantially affect any of the signals.

WD 1145+017 has a *Kepler*-band magnitude of $\mathrm{Kp} = 17.3$. Based on K2 observations of cool dwarf stars of similar brightness ($17 < \mathrm{Kp} < 17.6$), we expect a typical photometric precision of roughly 0.21% per 30 minute integration. After removing the six periodic signals from the K2 light curve, we find a typical scatter of 0.36% for the 30 minute integrations, indicating that photometry for WD 1145+017 appears to be somewhat noisier than typical stars of this brightness. This excess noise could be due to photometric noise intrinsic to the white dwarf, or it could be due to imperfect removal of the six signals we found.

**Periodogram Analysis** After processing the K2 data to remove systematics associated with the motion of the spacecraft, we performed a periodogram analysis. First, as part of a global search for periodic transits in K2 data, we calculated a Box Least Squared (BLS) periodogram[19]. Prior to calculating the BLS periodogram, we flattened the light curve with a basis spline to remove a long term (likely instrumental) trend. We then removed several points from the flattened light curve that were affected by strong cosmic rays by clipping 4-$\sigma$ upwards outliers. We calculated the BLS periodogram over periods ranging from 3.6 hours to 40 days and found a strong signal at a period of 4.499 hours. In addition to this 4.499 hour period (which we call the A period), the BLS periodogram showed five other strong periodicities in the light curve, which we label B through F.

We then calculated a Fourier Transform (FT) and a Lomb-Scargle (LS) periodogram of the corrected K2 data. Before calculating the LS periodogram, we flattened the light curve as described previously, which prevents low frequency power from leaking into the periodogram at high frequencies where thruster firing events have been removed[32]. Both the FT and the LS periodogram detect the same six periods found in the BLS as well as their harmonics. To better visualise the power at each period and its harmonics, we calculated a harmonic-summed LS periodogram[33] by adding the region of the periodogram near the fundamental 4.5-5 hour periods with the first two harmonics.

To confirm that the periodicities we found in periodograms of the corrected K2 light curve were not artifacts of our data processing, we performed an independent analysis of the light curve with a Systematics Insensitive Periodogram (SIP)[34]. The SIP is an alternate approach to mitigating systematic effects which takes advantage of the fact that the stars observed by K2 share common systematic modes. The SIP consists of a simultaneous fit of sinusoids and a number of eigenvectors (in this case, 150) determined from a principal component analysis of the light curves from all stars observed by *Kepler*. The SIP recovered the same periods and amplitudes as the LS periodogram, indicating that the periodicities are not artifacts.

We assessed the false alarm probability of the detections of these six periods using Monte Carlo techniques. We took the light curves from which we calculated the periodograms and randomly permuted the time series data many times. Each time, we recalculated the harmonic-summed LS periodogram, and recorded the peak of maximum power. After calculating the peak of maximum power for many different permutations, we found that the six peaks between 4.5 and 5 hours each have a $p < 10^{-4}$ probability of being spurious. We then sequentially removed each signal from the light curve and repeated the false alarm probability calculation. We found once again that each of the peaks had a false alarm probability of $p < 10^{-4}$. Finally, we assessed whether the periodicities we saw could be caused by incoherent red noise in the 4-5 hour range. We tested this by dividing the light curve into day long chunks, randomly permuting these chunks, and recalculating the harmonic summed periodograms many times. The resulting periodograms typically showed excess power in the range of 4-5 hours, but didn't show the strong peaks we see in the unscrambled data. When we estimated false alarm probabilities as before, we found once again, that each of the periodicities was significant with $p < 10^{-4}$.

We summarise the periodicities and their uncertainties in Table S2. We estimated uncertainties by calculating $\Delta P/P = P/(3\Delta t\sigma)$, where $\Delta P$ is the period uncertainty, $\Delta t$ is the time baseline of the K2 observations, $\sigma$ is the significance of each period detection, and 3 is the highest harmonic strongly detected in the K2 data. For the A-period, this estimate is roughly consistent with errors found from fitting the light curve with a transit model.

**Ground-Based Light Curves** We obtained ground-based photometric followup of WD 1145+017 in an attempt to detect the transits seen in the K2 data from the ground. We used the KeplerCam instrument on the 1.2-meter telescope at FLWO, and imagers on the MINERVA[35] and MEarth-South[36] telescopes. We observed WD 1145+017 with the FLWO 1.2-meter telescope in V-band on 23 March 2015, 25 March 2015, and 11 April 2015, and in R-band on 27 March 2015. We observed WD 1145+017 with four of the eight 0.4-meter MEarth-South telescopes on 17-19 April 2015, all using a 715 nm long pass filter, with the red end of the bandpass defined by a dropoff of the CCD quantum efficiency. Finally, we observed WD 1145+017 with one of the four MINERVA 0.7-meter telescopes on 18 April 2015 in white light, with both edges of the bandpass defined by the CCD quantum efficiency. MINERVA's field derotator was not enabled during the observations, which likely introduced systematics into the light curve. All of the ground-based follow-up light curves consisted of 1-minute integrations, so any sharp features in the light curves are artificially smeared by this amount of time. Even though the MEarth data has the same time-sampling as the FLWO and MINERVA data, the density of points is higher in the MEarth data because they come from four individual telescopes. We converted all of the data timestamps to BJD-TDB[37]. All of our ground-based observations are shown and summarised in Figure S1. As noted earlier, we observed two pairs of deep events on April 11 and 17. In addition, we see a possible single event on 23 March 2015 in FLWO data, and a possible event on April 18 at BJD = 2457130.7 seen simultaneously in MEarth-South and MINERVA data, with a depth of approximately 15% and in phase with the two MEarth-South events from the previous night. Most of our ground-based data have time baselines of over 5 hours, which means that if the deep transits occurred regularly on 4.5-4.9 hour periods without a changing depth, we would see at least one of them per 5 hours of observation. We do not see at least one deep transit in every night of observations, so we conclude that the depth of the transits must change significantly.

**Transit Analysis** We estimated the average depth and duration of the strongest transit signal seen in K2 by fitting the light curve with a solid-body transit model[38]. Even though the data are not consistent with solid body transits, this is a convenient model for measuring transit properties. The shapes of the light curve folded on the six periods identified in the K2 data are smeared by the 29.42 minute "long cadence" integrations, making the transit events appear shallower and longer than they are in reality. We accounted for the long cadence integration time of the K2 exposures by oversampling the model by a factor of 1000 and fitted the light curve using a multimodal nested sampling algorithm[39]. We found that when all 80 days of data were fitted simultaneously (assuming that the transit times do not vary), the K2 light curve was best described by longer duration (40-80 minute), shallow (1%-2%) transits. Under the assumption of strictly periodic transits, the behaviour seen in K2 is different from the 5 minute, 40% deep transits we observed in the ground-based data.

The K2 transits are not all inconsistent with the deep, short duration events we see in some of the ground-based light curves. There are two clusters (each lasting a couple of days) of deeper (approximately 4%) events in the K2 light curve that could be consistent with a short duration, deep event smeared out by the 30 minute *Kepler* long cadence exposure.

An interesting aspect of the transit fits is that the shape of the best-fit transit model is not consistent with solid objects transiting a white dwarf, and instead have much longer transit durations. In particular, the duration of the transits we inferred from the (highly smeared) K2 data and the durations we measured from the ground are longer than the expected transit duration of a small solid body across a white dwarf. The transit of a small body across a white dwarf in a 4.5 hour orbital period should take about a minute[40–42], while we see a 5 minute transit duration from the ground (see Figure S5) and a 40-80 minute duration in the K2 data. Equivalently, the stellar density we inferred from the K2 and ground-based transits is much lower than what we expect for a white dwarf. Such a discrepancy between known and inferred stellar density can be caused by several different effects[43]. We interpret the longer duration of the ground-based transits as evidence that the objects transiting the white dwarf are extended[44]. Given a transit duration $t_t$, the occulting region would have to have a size of roughly $(v_{orb} \times t_t - 2R_\star)$, which given our parameters is about 10 $R_\oplus$. The transits detected by K2 have an even longer duration, which could be an indication of unresolved transit timing variations[43] that would artificially smear out the phase-folded transit light curve.

We also placed limits on any solid body transits by analysing the first 15 days of the K2 light curve when the A-period was quiescent and not causing transits. We compared the mean flux value taken during a 30 minute window centered around the A-period's midtransit time with the mean flux value outside of this window. A 30 minute box shaped window is a good approximation for the effect of the 30 minute K2 exposure on a 1 minute transit. We found that at 3-$\sigma$ confidence, we can rule out one minute long transits with greater than 8% depth, equivalent to transits of a 0.4 $R_\oplus$ body, significantly larger than the bodies likely causing the transits we see here.

**Measuring Radial Velocities** We measured relative radial velocities from five individual MMT Blue Channel spectra using a cross correlation technique. The spectra were taken with a resolving power of $\Delta\lambda/\lambda = 4000$ over the wavelength range 356 nm to 451 nm, with a signal to noise ratio of about 25 per resolution element. We first confirmed that the relative shifts in the spectra were significantly smaller than one pixel by fitting the calcium K line at 393.4 nm for each spectrum with a Gaussian line profile, and then constructed a template spectrum by summing the five individual exposures. We then calculated the cross correlation function between the coadded template spectrum and the individual spectra, and took its peak to be the relative shift of the individual spectra. We converted this shift into a relative radial velocity, corrected for the Earth's barycentric motion, and found that the radial velocity of WD 1145+017 is stable at the level of 500 m s$^{-1}$.

We then calculated upper limits on the mass of a companion in a close orbit by fitting the RV measurements with a circular Keplerian orbit. We held the orbital period constant at the 4.5 hour period of the strongest transit signal, and allowed the time of transit to float freely (given the uncertainty in the transit times). We fitted the RV data using a Markov Chain Monte Carlo technique with an affine invariant sampler[45], and found that the data are consistent with no variations and strongly exclude stellar mass companions ($m > 0.08 M_\odot$). We exclude companions with masses greater than 10 Jupiter masses at 95% confidence. This limit is weaker than the 500 m s$^{-1}$ RV stability would imply because the time sampling of our radial velocity observations (taken at roughly 2 hour intervals) permitted higher semi-amplitude RV solutions to be fit with our measurements taking place when the fitted curves are near the systemic velocity. To be conservative, we have reported this weaker limit to massive companions. When we fitted the RV data while fixing the ephemeris to the FLWO and MEarth data, we found that objects more massive than 1.5 Jupiter masses are excluded at the 95% confidence level. We also fitted the RV data while allowing the ephemeris to float and allowing the orbital period to vary between 3 and 6 hours, and found that we could exclude companions with masses greater than 11 Jupiter masses with 95% confidence.

**Adaptive Optics Imaging and Background Object Exclusion** We obtained near infrared high resolution adaptive optics (AO) imaging in an attempt to rule out contamination of the K2 aperture by background objects that could potentially dilute the transits or be the source of the observed transits in the K2 or ground-based light curves. We observed WD 1145+017 with the wide detector of the NIRC2 instrument behind the Keck II Natural Guide Star Adaptive Optics (NGSAO) system on the Keck II telescope on 7 April 2015. We observed for 270 seconds (composed of three 90 second dithered sub-exposures) in both J and Ks bands using WD 1145+017 as the guide star.

Due to relatively poor observing conditions and the faintness of the guide star, AO performance was suboptimal. Nevertheless, we managed to achieve AO lock and detected no sources within 12 arcseconds of WD 1145+017. The data exclude stars at 5-$\sigma$ confidence that are within four magnitudes of the primary target and outside of 1 arcsecond, and within 2.5 magnitudes at a separation of 0.5 arcseconds. The 5-$\sigma$ sensitivities were estimated by injecting and recovering sources into the frame at various separations and azimuthal positions from the primary target. In particular, we experimented with injection of nearly equal brightness companions to WD 1145+017 (0.5 magnitudes fainter), and found that the AO images are sensitive to such sources at an angular separation of 0.1 arcseconds. This limit is equivalent to a projected physical separation of 20 AU.

In Figure S2, we show the AO images and archival images which exclude contaminating stars from the K2 photometric aperture. WD 1145+017 has a total proper motion of about 43 milli-arcseconds per year[46], meaning that its position in the sky has moved by about 3 arcseconds with respect to background sources since it was imaged in the National Geographic Society-Palomar Observatory Sky Survey (POSS I) in 1952. If there were a background star bright enough to cause the transits we see at the current position of WD 1145+017 we would see an an elongated source in the POSS image. Instead, the POSS image is consistent with a single source.

The imaging presented here makes it possible to rule out many scenarios in which the transits are due to photometric variability in background objects. Additional constraints on blend scenarios come from the 5 minute duration and 40% depth of the transits. Such a deep and short duration transit must be from a very dense star of nearly the same brightness as the foreground white dwarf, but we see no evidence in photometry or spectroscopy for any such star with a different temperature. A nearby companion white dwarf with the same surface temperature could be the basis for a plausible background blend scenario, but this would not change our conclusion that we observed transits of a white dwarf.

**Measuring Temperature and Elemental Abundances** We measured the temperature and elemental abundances of WD 1145+017 using photometry from the Sloan Digital Sky Survey (SDSS)[47] and the MMT spectra. We first determined the effective temperature of WD 1145+017 by fitting the SDSS photometry using a grid of helium atmosphere models that include traces of metals close to our final adopted abundances. We accounted for interstellar reddening using an iterative approach[48] and found a value of $A_g = 0.016$. We measure an effective temperature of $T_{\rm eff} = 15{,}900 \pm 500$ K. This is slightly cooler than the value determined in a previous study[49] which measured $T_{\rm eff} = 16{,}900 \pm 250$ K using pure helium atmosphere models. Had we used pure helium models, our fit to the photometry would have yielded $T_{\rm eff} = 16{,}700$ K, in agreement with the previously determined spectroscopic values[49].

Since the presence of heavy elements and hydrogen, the abundances of which cannot be well constrained from existing data, can significantly affect the surface gravity determination, we keep $\log g_{\rm cgs}$ fixed at a value of 8.0, which is typical for white dwarfs[50]. This corresponds to a white dwarf with a mass of 0.6 $M_\odot$ and radius of 1.4 $R_\oplus$, which we adopt in this paper. Using these atmospheric parameters, we then calculated grids for each heavy element and fitted the MMT spectroscopic observations. We followed a method used in previous studies[51,52] and iteratively fit several small portions of the spectrum to obtain a given metal abundance, while keeping the abundances of all other elements fixed to the values found in the previous iteration. Unfortunately, because the spectrograph's resolution element is one angstrom, we cannot resolve some lines and it is difficult to determine accurate abundances. We identify features from 6 heavy elements (magnesium, aluminium, silicon, calcium, iron and nickel) and possibly titanium and chromium as well (see Figure S3). Using the non-detection of the hydrogen gamma line, we place an upper limit on the abundance of hydrogen at [H/He] $\lesssim -4.5$. In the future, high resolution spectroscopy should confirm the presence of these elements and give precise measurements of their abundances.

Finally, we *a posteriori* calculated individual models with temperatures increased and decreased by 1,000 K and surface gravity increased and decreased by 0.25 dex. These models are unable to describe the neutral helium lines as well as models using our adopted spectroscopic parameters.

We calculated the cooling age of WD 1145+017 using evolutionary models[53] modified for helium envelope white dwarfs with layers of hydrogen and helium with typical mass fractions $q(He) = 10^{-2}$ and $q(H) = 10^{-10}$. Assuming conservative errors of 0.25 dex on surface gravity, we find that WD 1145+017 has a cooling age of $175 \pm 75$ Myr.

**Measuring Infrared Excess** After determining the best-fit white dwarf model from SDSS optical photometry and the MMT spectra, we examined near infrared photometric measurements from United Kingdom Infrared Telescope (UKIRT) Infrared Deep Sky Survey (UKIDSS)[54] and Wide-field Infrared Survey Explorer (WISE)[55] images of WD 1145+017. We found that near infrared photometry from UKIDSS and WISE show evidence for excess infrared emission. We quantified this excess emission by fitting photometric measurements of WD 1145+017 with a metal-rich white dwarf model (fixed to the effective temperature and elemental abundances we found previously) and a second blackbody emission component, taking interstellar reddening into account. We fitted this model to the data using a Markov Chain Monte Carlo technique with an affine invariant sampler[45,56].

We found that the photometry at optical wavelengths is well described by the white dwarf model spectrum, however, near infrared photometric measurements favour the presence of a second blackbody component with an improvement in $\chi^2$ of 39 (for 11 datapoints with the addition of two model parameters) over a model without excess infrared emission. The reduced $\chi^2$ (or $\chi^2$ per degree of freedom) improves from 5.3 to 1.5 with the addition of the infrared excess into the model. The second blackbody has an effective temperature of $1,145^{+240}_{-195}$ K, and has a projected area $116^{+129}_{-61}$ times greater than that of the white dwarf. We interpret this excess emission as evidence for a warm debris disk near the Roche radius of the star[8,9,12,16] for rocky bodies. Our best–fitting model is shown in Figure S4. We then fitted a dusty ring model[9] to the photometric measurements, but found that the existing photometric data are insufficient to constrain the model. Another possibility is that the IR excess is caused by a nearby brown dwarf companion, though this scenario is far less common than dusty disks. Additional IR photometry or spectroscopy can distinguish these two explanations.

In the case of WD 1145+017 because we see transits, it is likely that the debris disk has an inclination angle close to 90 degrees. If this is the case, the actual luminosity of the disk is likely much larger than we see due to projection effects. The model parameters we find are consistent with the parameters derived for other white dwarf infrared excesses when high inclination angles are assumed[14,16]. If the disk is flat and within a few degrees of $i = 90$ degrees, then the projected area of the disk would be small, and we would not detect its emission. Our detection of an IR excess implies that the disk is either out of the plane of the transiting objects or is warped or flared. Alternatively, the disk could be unrelated to the debris we see transiting, implying that the system could be very dynamically active[17,57].

**Estimate of Mass Loss** We estimated the amount of mass loss necessary to explain the transits of WD 1145+017 with a dusty tail. We consider a plane of material passing in front of the white dwarf with a surface mass density $\Sigma$, and a height $h$ equal to the diameter of the white dwarf ($d_\star = 2R_\star$) moving with velocity $v_{rel}$ relative to the planetesimal across the line of sight. The mass-loss rate $\dot{M}$ from the planetesimal is given by:

$$\dot{M} \approx \Sigma\, v_{\rm rel}\, h\,. \qquad (1)$$

We estimate that the velocity of the dust tail with respect to the motion of the planetesimal, $v_{rel}$, is in the range between the escape speed, $v_{\rm esc}$, from the planetesimal and $2\beta v_{orb}$, where $\beta$ is the ratio of radiation pressure to gravity[21], and $v_{orb}$ is the orbital velocity of the planetesimal (i.e., 0.5 $\lesssim v_{\rm rel} \lesssim 15$ km s$^{-1}$). The appropriate value of $v_{\rm rel}$ to use depends on the details of how the escaping atmosphere condenses into dust grains and then joins the radiation-pressure dominated part of the flow. The average depth of the transits in the K2 light curve is about 1%, but this depth is diluted by the 30 minute K2 integration times. From the ground-based data, we know that the typical (full width at half maximum) of the transits is about 4 minutes, so the average undiluted transit likely has a depth of about 8%, which we adopt as our value for optical thickness of the attenuating sheet of dust, $\delta$ (assumed $< 1$). In what follows, we assume that the Mie scattering cross section of the dust particles is roughly equal to their geometrical cross section (a good approximation for particles $\gtrsim 1/3$ micron in size). If the dust particles have a radius $a_g$, then the surface density $\Sigma \approx \rho a_g \delta$ and we can write:

$$\dot{M} \approx \rho \, v_{\rm rel} \, a_g \, d_\star \, \delta \qquad (2)$$

where $\rho$ is the material density of the grains. Adopting $\rho \simeq 3$ g cm$^{-3}$, $v_{\rm rel} \simeq 3$ km s$^{-1}$, $a_g \simeq 1/2$ $\mu$m, $d_\star \simeq 2.8$ R$_\oplus$, and $\delta \sim 0.1$, we estimate $\dot{M} \sim 8 \times 10^9$ g s$^{-1}$ with an uncertainty of about a factor of 4.

**Tidal Locking of the Planetesimals** We show below that essentially any minor planets in a 4.5 hour orbit around a white dwarf will have rotation periods tidally locked to their orbital periods. We consider a minor planet with the mass and radius of Ceres. The timescale for tidal locking $t_{\rm circ}$ for the planetesimals orbiting WD 1145+017 is given as[58]:

$$\Delta t_{\rm circ} \sim \frac{8 m_p Q'_p a^6}{45 G M_\star^2 r_p^3} \Delta \omega \qquad (3)$$

where $m_p$ is the mass of the planetesimal, $r_p$ is the radius of the planetesimal, $a$ is the orbital semi-major axis, $M_\star$ is the stellar mass, $G$ is the gravitational constant, $\Delta \omega$ is the change in rotational frequency between the body's initial rotation and its final tidally locked rotation, and $Q'_p$ is a constant describing tidal dissipation, given by:

$$Q'_p = Q_p \left( 1 + \frac{38 \pi r_p^4 \gamma}{3 G m_p^2} \right) \qquad (4)$$

where $Q_p$ is the tidal dissipation function and $\gamma$ is rigidity of the material. We take $Q_p \sim 100$ and $\gamma \sim 7 \times 10^{11}$ dyne cm$^{-2}$, values appropriate for rocky solar system bodies[58], which yield $Q'_p \sim 2.5 \times 10^5$. Using this and values for Ceres in Equation 3, we find $t_{\rm circ} \sim 15$ years. Thus, we can safely say that any bodies in 4.5 hour orbits almost certainly have become tidally locked.

**Mass Loss from a Thermal Wind** Since any Ceres sized body in a 4.5 hour orbit around a white dwarf will be tidally locked, then heat redistribution will be suppressed, and the temperature at the substellar point on the planetesimal will be:

$$T_{\rm eq,ss} \simeq \sqrt{\frac{R_\star}{a}}\, T_{\rm eff} \qquad (5)$$

Plugging in stellar and orbital parameters gives $T_{\rm eq,ss} \simeq 1{,}675$ K. The mean equilibrium temperature over the facing hemisphere, $\langle T_{\rm eq} \rangle$, is

$$\langle T_{\rm eq} \rangle \simeq 2^{-1/4} \sqrt{\frac{R_\star}{a}}\, T_{\rm eff} \qquad (6)$$

which comes out to be 1,410 K.

We can now calculate the vapour pressure $p_{\rm vap}$ of materials at the surface of the planetesimal using:

$$p_{\rm vap} = \exp\left[-\frac{mL_{\rm sub}}{k_B T_{\rm eq}} + b\right] \qquad (7)$$

where $m$ is the molecular mass of the material, $L_{\rm sub}$ is the latent heat of sublimation, $k_B$ is Boltzmann's constant, and $b$ is an empirically measured constant. We calculated vapour pressures for a variety of minerals using values taken from a previous study[59] or using values we calculated with the MAGMA code[60,61]. The results are shown in the top panel of Fig. S6 as a function of temperature. Vapour pressures for orthoclase, albite, iron, and fayalite are in the range from $0.01 - 10$ dyne cm$^{-2}$ at the relevant temperatures.

We computed the Jeans escape parameter, $\lambda = (v_{\rm esc}/v_{\rm therm})^2$, for each mineral on the surface of the fiducial planetesimal, where $v_{\rm esc}$ is the escape velocity and $v_{\rm therm}$ is defined as $\sqrt{2k_B T/m}$. Once these materials have sublimated, they typically dissociate into oxygen ($O_2$) and metal oxides, so the typical molecular mass is of order 30-100 amu. It turns out that for essentially all of the minerals we considered, $\lambda \lesssim 1$. Thus, we take the mass loss from the planetesimal to be in the free streaming limit.

We then calculate the mass-loss flux $J$ from:

$$J = \alpha\, P_{\rm vap} \sqrt{\frac{m}{2\pi k_B T}} \qquad (8)$$

where $\alpha$ is the 'sticking coefficient' and is typically of the order of $0.1 - 0.3$. We then multiply $J$ times the area of the heated hemisphere of the planetesimal to yield the total mass-loss rates. The results are shown in panel **b** of Figure S6 as a function of temperature. The grey shaded region denotes the range of mass-loss rates that could plausibly account for the observed transits if they are due to dust that condenses from the thermal winds the planetesimal releases. We find that the minerals listed above could sublimate and be lost from the planetesimal at the rates we infer from the transits.

**Dynamical Stability** We estimated the conditions required for long term orbital stability in two ways. First, we computed the fractional orbital separations inferred for the different planetesimals, and compared them to the Hill radius for each body. We considered a set of bodies in concentric, circular orbits, each with radius $a_n$, around a central object of mass $M_\star$. The orbits must be circular, or at least very close to circular, because of their tight spacing. Orbits which cross one another will be unstable on short timescales[17] or lead to collisions. For non-crossing orbits, we require:

$$\frac{\Delta a_n}{a_n} = \frac{2}{3}\frac{\Delta P_n}{P_n} > 1 - \frac{d_{\rm peri}}{a_n} = e \qquad (9)$$

where $d_{\rm peri}$ is the periastron distance of one of the planetesimal's orbits, and $e$ is the orbital eccentricity. For period spacings of roughly 1%, the eccentricity of the orbits must be less than roughly 0.007 to prevent crossing orbits (unless the arguments of periastron are aligned for all of the planetesimals, which is unlikely).

To calculate the Hill radius, for simplicity, we assumed that all of the planeteismals have the same mass, $m_p$. The fractional orbital separation can be expressed both in terms of the fractional differences in the orbital periods, $P_n$, and in terms of the Hill sphere radii, $R_h$:

$$\frac{\Delta a_n}{a_n} = \frac{2}{3} \frac{\Delta P_n}{P_n} = \xi \frac{R_h}{a} = \xi \left(\frac{m_p}{3M_\star}\right)^{1/3} \quad (10)$$

where $\xi$ is the number of such Hill radii that separate orbit $n$ from $n+1$. For stability we required

$$m_p \lesssim \frac{8}{9} \frac{1}{\xi^3} \left(\frac{\Delta P_n}{P_n}\right)^3 M_\star \quad (11)$$

The closest period spacings in WD 1145+017 range from 0.7% to 1.4%. If we use Eqn. (11) with $M_\star$ set equal to 0.6 $M_\odot$, we find:

$$m_p \lesssim \frac{0.06 \text{ to } 0.5}{\xi^3} M_\oplus \quad (12)$$

For long term stability[62], $\xi$ should be $\sim$10. This yields $m_p \lesssim (0.6 \text{ to } 5) \times 10^{-4} M_\oplus$, which includes Solar systems bodies such as Vesta, Ceres, Charon, and Haumea.

We also tested for the stability of the orbits using numerical N-body integrations. We used a fourth-order Runge-Kutta integration scheme to simulate the N-body interactions of six planetesimals orbiting the white dwarf. The starting conditions were coplanar circular orbits with semi-major axes corresponding to the periods found in the K2 light curve. We performed integrations over $2 \times 10^6$ orbits (1000 years) with planetesimals of varying masses. We found that in such close orbits, planetesimals with masses comparable to that of Earth's moon significantly perturbed the other planetesimals' orbits in such a way as to quickly excite instabilities. Slightly smaller planetesimals with masses comparable to those of Ceres or other large asteroids in the Solar System remained in stable orbits over the timescale of our simulations.

**Dust Tail Simulations** We simulated the dynamics of dust particles after being launched from a planetesimal in a 4.5 hour orbit around WD 1145+017. We initialised the simulation by launching dust in a ± 60 degree cone pointed towards the host star from the Roche radius of the planetesimal at 1.5 times the planetesimal's escape velocity. The dust particles were drawn from a power law distribution with a slope of -2, a maximum particle size of 1 micron, and a minimum size of 0.05 micron. We computed dust particle trajectories in the rotating frame in which the host star and planetesimal are fixed. We accounted for both the star and planetesimal's gravitational force and radiation pressure from the white dwarf's luminosity using the Mie scattering cross section for the dust particles. Dust particles with generic indices of refraction $n = 1.6$ and $k = 0.01$ were used.

We found that even though the intensity of the radiation field is substantially lower than that of other disintegrating planets (due to the white dwarf's intrinsically low luminosity), the dust particles still formed a well-defined trailing dust tail, consistent with the asymmetric transits seen in the ground-based data. The dust also formed into a weaker leading tail. We show the results of our simulations in Figure S7. The vertical extent of the dust tail shown here is slightly smaller than we would expect given 40% transit depths, which might indicate that higher particle launch velocities than we used are appropriate, but the simulation input parameters are too uncertain to say for sure.

**Occurrence of Disintegrating Minor Planets around White Dwarfs** Despite there being only one known instance of a disintegrating minor planet transiting a white dwarf, we provide a rough estimate of the occurrence rate of such objects. Previous work searching for planets transiting white dwarfs has yielded upper limits[40,63] (these non-detections of planets are consistent with N-body simulations[64]), but these searches were focused on solid-body transits, and might not have been sensitive to objects with such highly variable transit depths. WD 1145+017 is one of 161 white dwarfs which were observed during Campaigns 0, 1, and 2 of the K2 mission, as part of programmes to search for transits and eclipses. These observing programmes included most known white dwarfs in the three K2 fields. After searching these 161 targets, our transit search pipeline identified 7 objects whose light curves showed "threshold crossing events" where the BLS periodogram indicated a significant peak. Four of these events were caused by data artifacts, two of the objects are likely eclipsing binary stars, and the final one is WD 1145+017.

The transit probability of a body is given as $p_{\text{transit}} = (R_t + R_\star)/a$, where $R_t$ is the size of the transiting object, in our case a dust tail. In order to give a 40% transit depth, the vertical extent of the dust tail is likely of order the size of the white dwarf, so $p \simeq 2R_\star/a$. Plugging in appropriate values yields $p_{\text{transit}} \simeq 0.02$. If we assume that our transit detection pipeline has close to 100% completeness (a good assumption for high significance events like the transits of WD 1145+017), then the occurrence rate of disintegrating planets around white dwarfs is about 0.3. We estimated uncertainties by calculating the relative probability of detecting one disintegrating planet around 161 white dwarfs with a binomial probability distribution given different underlying occurrence rates, while imposing a Jeffreys prior on occurrence rate. The 68% confidence interval is { 0.05, 0.5}, and the 95% confidence interval is { 0.007, 0.85}. We show the resulting probability distribution in Figure S9.

**Debris Clouds as the Source of the Transits** Finally, we consider here the possibility that the transits are caused by debris clouds from a disk filling the Hill spheres of passing asteroids that have been broken (or, dislodged) from larger rubble piles by collisions[27]. We start by computing the Hill sphere radii for individual asteroids or planetesimals of sizes ranging from 1 km to 1000 km, each assumed to have a density of 4 g cm$^{-3}$, as a function of orbital distance from a 0.6 $M_\odot$ white dwarf. We assume the orbits are circular for simplicity, though the problems only become more restrictive with eccentric orbits. The Hill sphere radii of these planetesimals are shown in Figure S8, where each of the six blue curves is labelled with the planetesimal's radius. We also show on the same plot the radii of opaque spheres that can produce transits of different durations as they transit the white dwarf. The six red curves for constant transit durations are labelled by the transit time in minutes. The vertical grey shaded region marks the location of an orbit with a 4.5 hour period. There are no curves with transit durations of 5 minutes or greater (as observed from the ground-based transits) that intersect the Hill sphere radii curves for planetesimals smaller than 1000 km with orbital separations smaller than about $10 R_\odot$. Such intersection points correspond to orbital periods longer than 100 days.

For orbital periods near 5 hours, orbiting bodies are either already filling their Roche lobes or are under-filling them by at most a factor of about 2. Therefore, there is not much available volume for the planetesimals to carry along a sizable debris cloud. The spreading chains of asteroids that are discussed in a previous study[27] which come from a disrupted rubble pile would have sizes smaller than a kilometre, in which case $\sim 10^8$ of them would be necessary to block a significant fraction of the light of the white dwarf – and only then if all the debris were concentrated over a tiny fraction of the orbit. One could also imagine a 1000-km planetesimal breaking up into $10^3$ smaller (100 km) bodies. The debris clouds around any of these could block at most ~0.1% of the white dwarf's light. Then, again, one would need all 1000 of these planetesimals collecting along a small arc of the orbit to block the light of the white dwarf. According to calculations in the previous study[27], such concentrations would not persist for the roughly 400 orbital cycles observed by K2.

Trailing dust tails avoid this problem because they are not gravitationally bound to the bodies that emit them, and are instead constantly replenished. We conclude that it is easier to explain the transits with dust which is not bound to the bodies that emit it.

Table S1: Summary of parameters for WD 1145+017.

| Parameter | Value | Uncertainty | Source |
|---|---|---|---|
| **Spectroscopic parameters** | | | |
| Stellar effective temperature | 15,900 K | ± 500 K | this work |
| **Model derived parameters** | | | |
| Distance to star | 174 pc | ± 25 pc | this work |
| White dwarf cooling age | 175 Myr | ± 75 Myr | this work |
| **Astrometric parameters** | | | |
| Right Ascension | 11:48:33.627 | | SDSS |
| Declination | +01:28:59.41 | | SDSS |
| Right ascension proper motion | -43.3 mas/yr | ± 4.9 mas/yr | Gaia Input |
| Declination proper motion | -7 mas/yr | ± 4.9 mas/yr | Gaia Input |
| **Photometric measurements** | | | |
| u | 16.969 | ± 0.008 | SDSS |
| g | 17.029 | ± 0.004 | SDSS |
| r | 17.380 | ± 0.005 | SDSS |
| i | 17.608 | ± 0.006 | SDSS |
| z | 17.840 | ± 0.018 | SDSS |
| Y | 17.434 | ± 0.019 | UKIRT |
| J | 17.504 | ± 0.028 | UKIRT |
| H | 17.529 | ± 0.059 | UKIRT |
| K | 17.397 | ± 0.081 | UKIRT |
| W1 | 17.023 | ± 0.159 | WISE |
| W2 | 16.507 | ± 0.351 | WISE |

Table S2: Periods detected in K2 data. Transit times are given in BJD - 2454833, and correspond to the mid-event time shown in Figure 1. For the weaker periods (C-F), the midtransit times are poorly defined.

| Name | Period | Uncertainty | Transit Time |
|---|---|---|---|
| A | 4.49888 hours | ± 0.25 seconds | 1977.307334 |
| B | 4.60530 hours | ± 0.53 seconds | 1977.178422 |
| C | 4.78283 hours | ± 1.43 seconds | 1977.206917 |
| D | 4.55000 hours | ± 1.29 seconds | 1977.120758 |
| E | 4.82336 hours | ± 1.45 seconds | 1977.175618 |
| F | 4.85848 hours | ± 1.47 seconds | 1977.240058 |

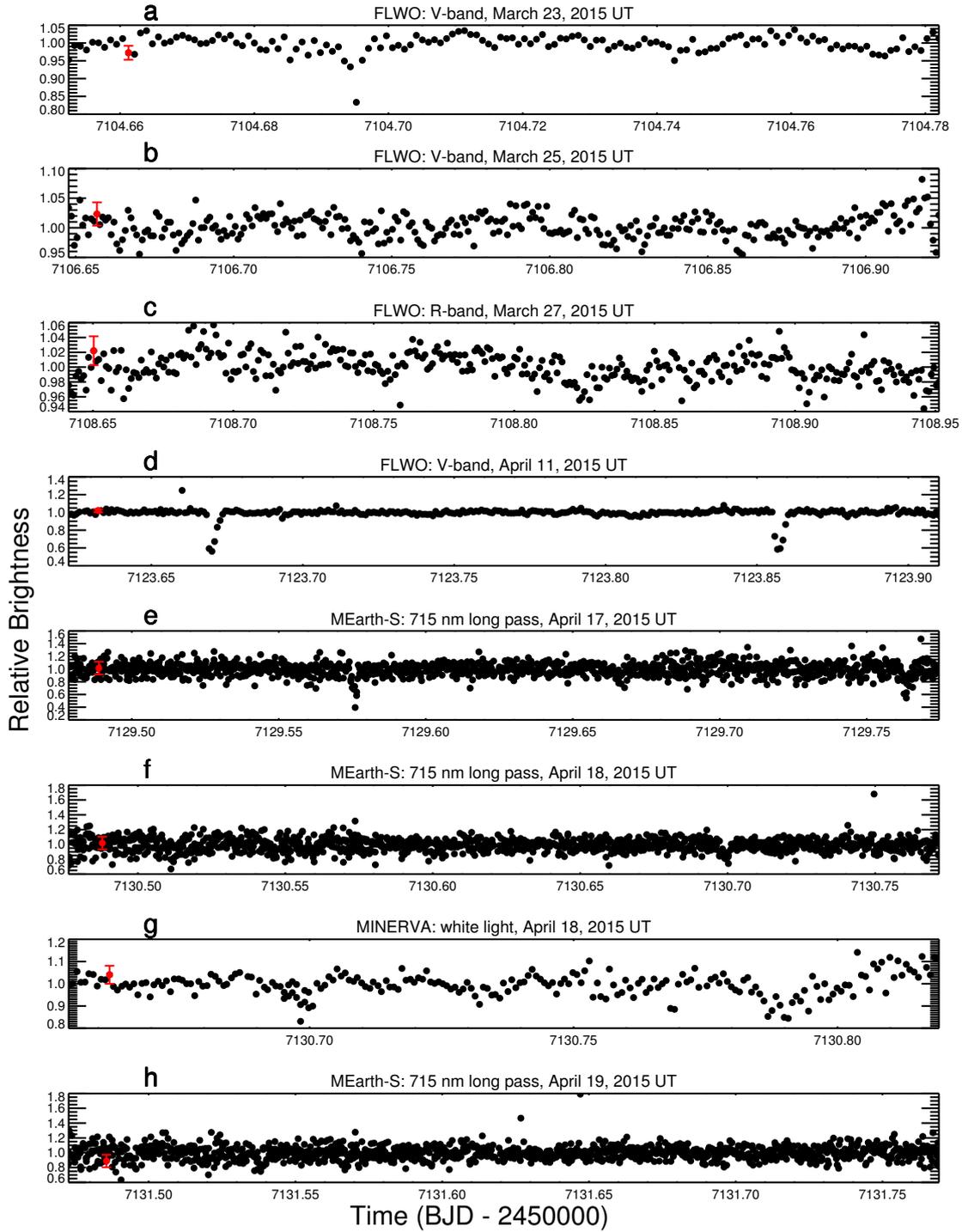

Figure S1: All ground-based light curves obtained between 23 March 2015 and 19 April 2015. We detected transits 4.5 hours apart on April 11 and April 17, possible single events on March 23 and April 18, and no convincing signals on three other nights. We show the typical measurement uncertainty (standard deviation) with a red error bar on one datapoint in each panel. Note the differences in y scale between the panels.

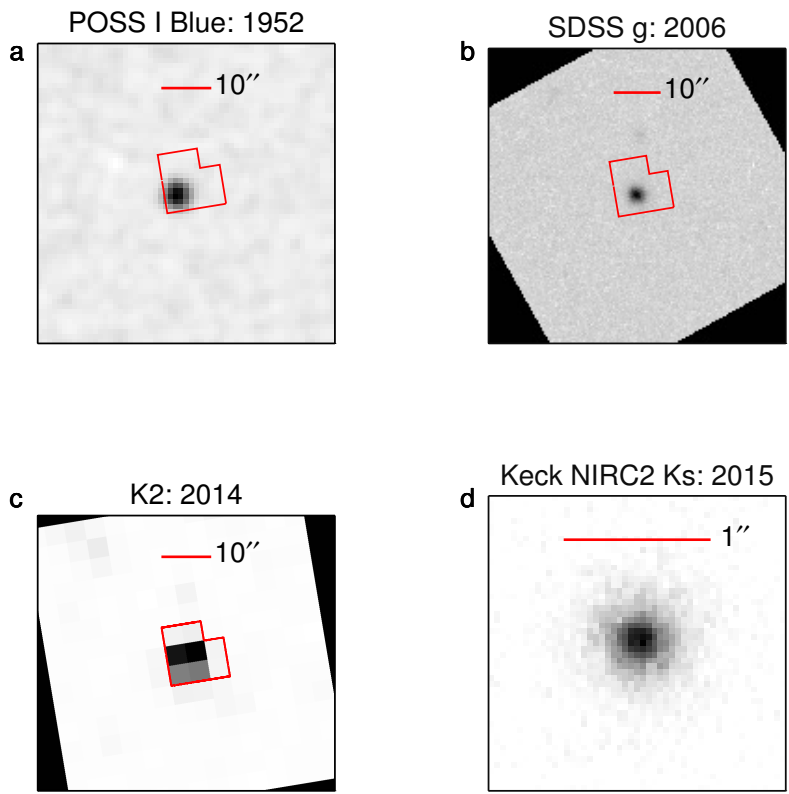

Figure S2: Images of WD 1145+017. The image scale is given in arcseconds. Panel a: A POSS-I image from 1952 with a blue-sensitive photographic plate. Panel b: An SDSS image from 2006 in a green filter. Panel c: A coadded image taken by K2 during its 2014 observations. Panel d: A high resolution adaptive optics image taken at Keck Observatory in 2015 in near infrared wavelengths near 2 microns. These images show no evidence for any other nearby stars bright enough to be the source of the transits. In panels a-c, the K2 photometric aperture is overlaid in red.

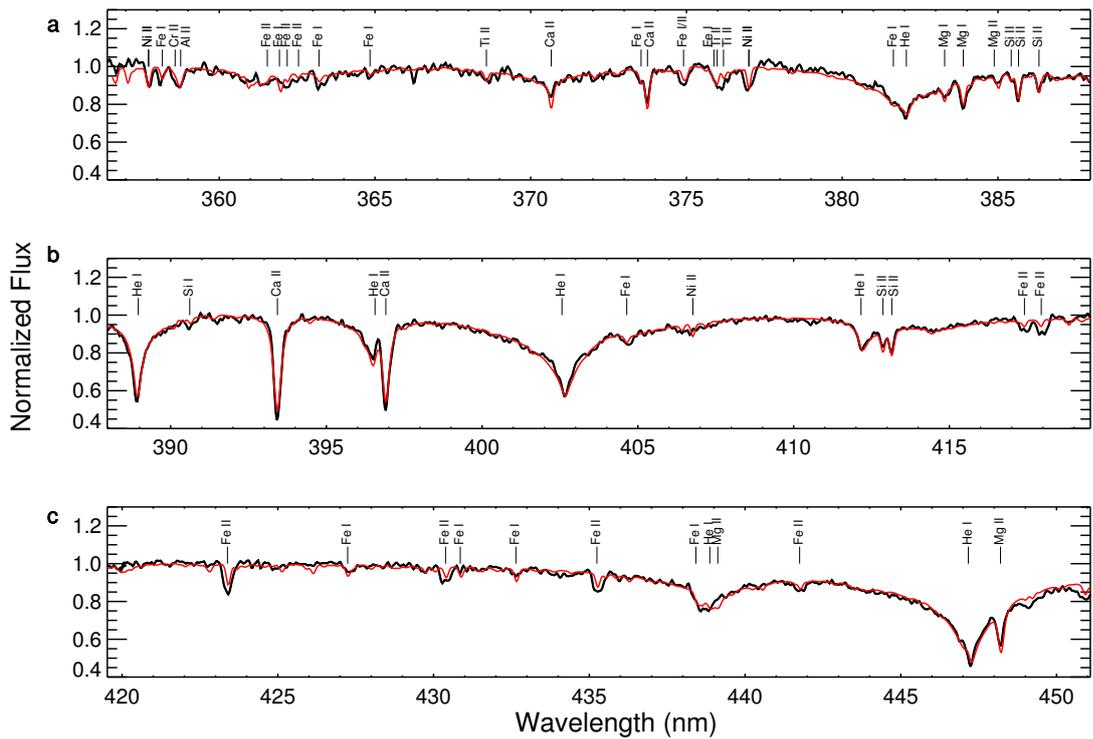

Figure S3: Spectrum of WD 1145+017 and best-fit model. We obtained five 300 second exposures of WD 1145+017 with the MMT Blue Channel spectrograph, summed them together, continuum normalised, and smoothed over a resolution element to produce this figure. We overplot a model spectrum in red, which we fit to the data to measure elemental abundances. The spectrum shows absorption features corresponding to magnesium, aluminium, silicon, calcium, iron, and nickel, which we label. We also see some evidence for titanium and chromium lines.

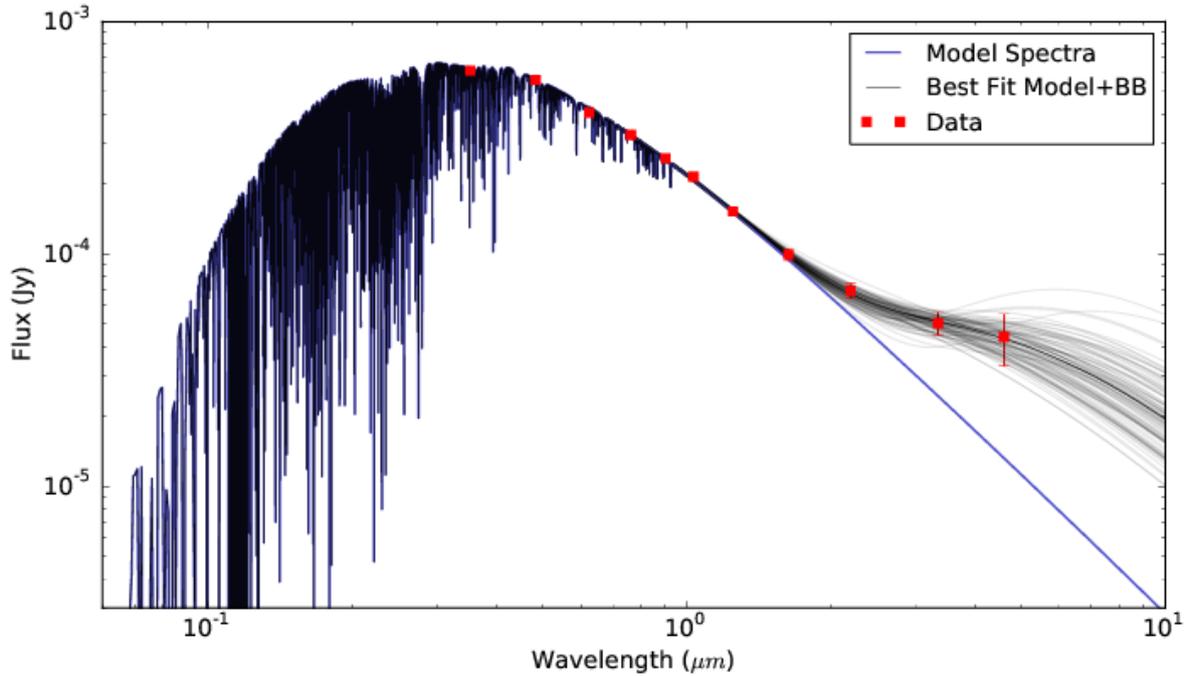

Figure S4: Fit to photometric measurements of WD 1145+017. Photometric data are shown in red, the white dwarf model spectrum is shown in blue, and the combined white dwarf + blackbody model is shown in black. The units of flux are janskys ($10^{-23}$ ergs s$^{-1}$ cm$^{-2}$ Hz$^{-1}$). Transparent grey lines show random draws from our posterior probability distribution. We took photometric measurements from SDSS, UKIDSS, and the WISE mission. We found evidence for excess infrared emission, which we modelled with a blackbody curve. The inferred infrared excess is consistent with those found around other white dwarfs. Error bars shown are the 68% confidence interval for each point.

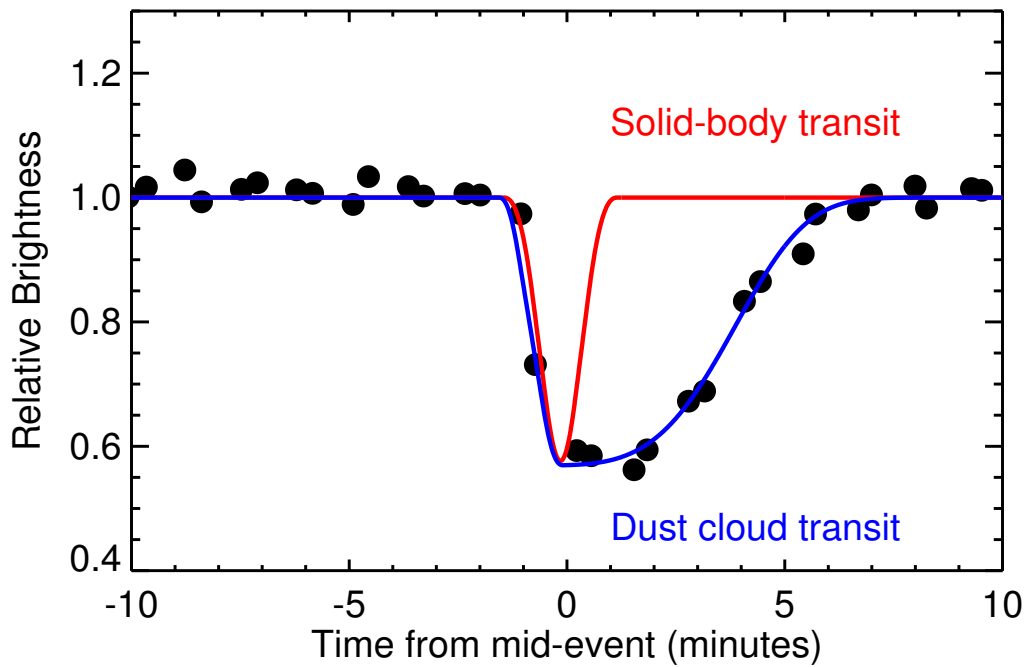

Figure S5: FLWO transits compared with a solid body transit model and a dusty tail transit model. The solid body model (shown in red) is calculated for a sub-Earth sized planet transiting a white dwarf with our stellar parameters in a 4.5 hour orbit, and has a much shorter duration than we see. The dust transit model (blue) simulates a dusty tail crossing over the star with optical depth $\tau \propto \exp(-ax^4)$, where $a$ is a free parameter. The typical measurement uncertainties are smaller than the size of the symbols.

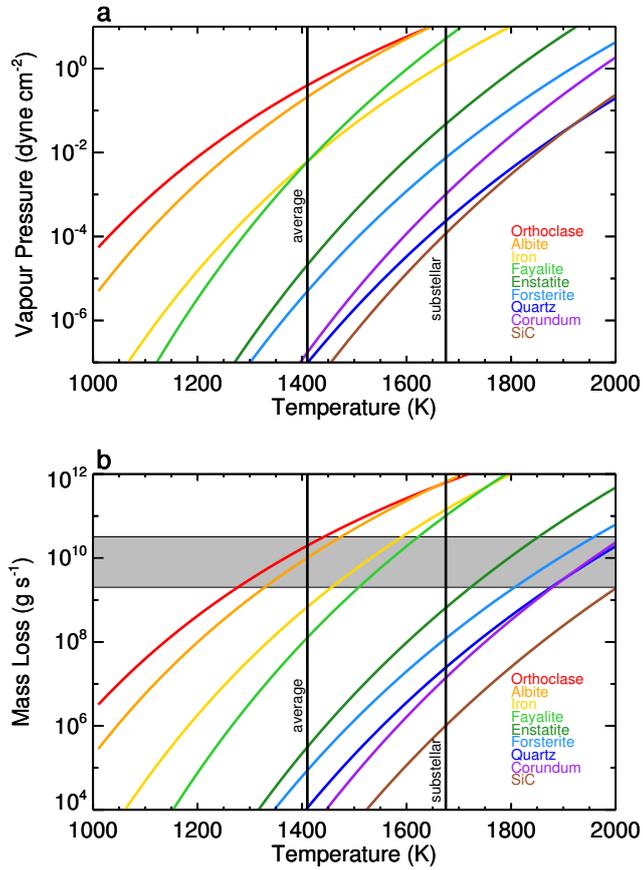

Figure S6: Vapour pressure and mass-loss rates for various refractory materials. Panel a shows calculated vapour pressure versus the surface temperature of the planetesimal, and panel b shows the mass-loss rate from a Ceres-like planetesimal as a function of surface temperature. For the mass-loss rate we assume free streaming from the planetesimal because the Jeans escape parameter, $\lambda$, is close to unity. The planetesimal is taken to have the size and mass of Ceres. The vertical solid lines indicate the sub-stellar equilibrium temperature and the average temperature of the facing hemisphere. The grey shaded region in the mass-loss plot indicates the range required to explain the transits.

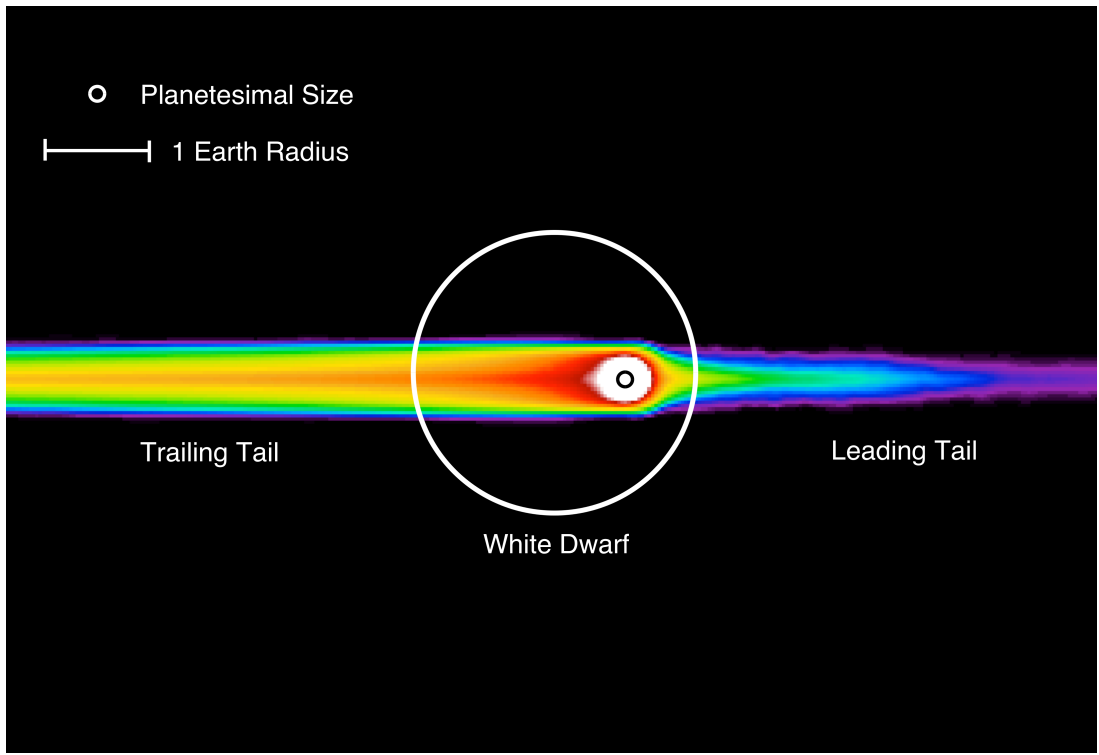

Figure S7: Simulation of a dust cloud produced by a disintegrating planetesimal orbiting WD 1145+017. For reference, the size of the rocky planetesimal is shown in the top left. The dust forms into a relatively strong trailing tail and a weaker leading tail.

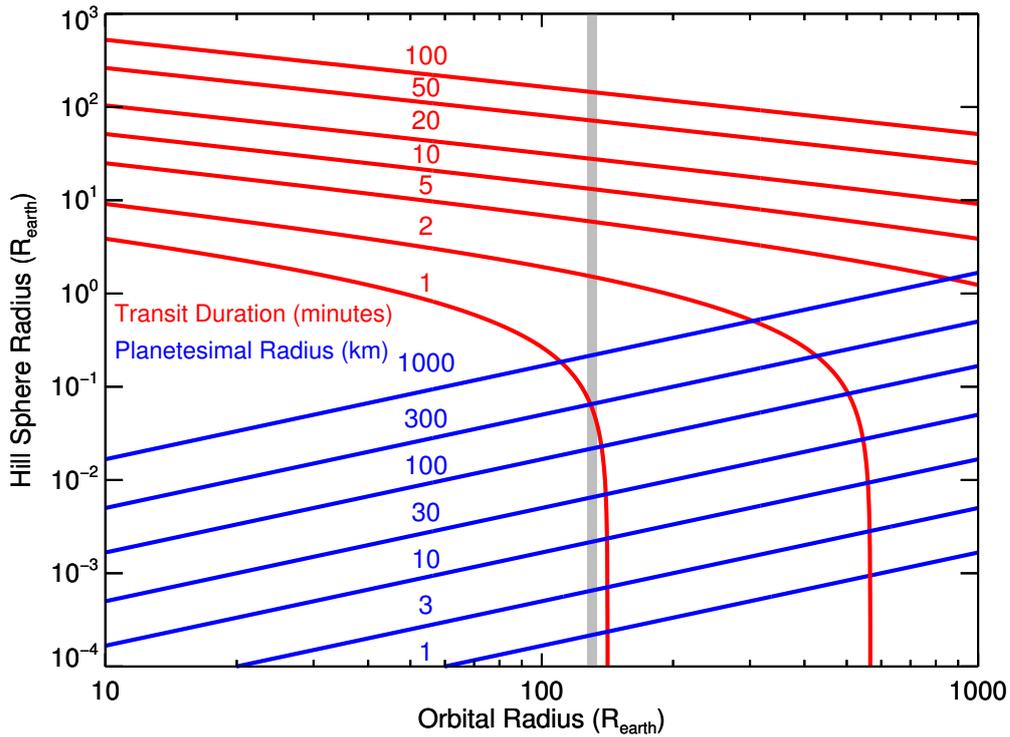

Figure S8: Hill sphere radii of various planetesimals versus orbital radius. Red lines represent the size of the Hill sphere necessary to produce a transit with various durations, labelled in red near the curves. The blue lines represent the size of the Hill sphere for rocky planetesimals of varying sizes. Long lived gravitationally bound debris clouds surrounding planetesimals in the mass range we consider can't produce transit durations longer than roughly one minute in a 4.5 hour orbit (grey shaded region).

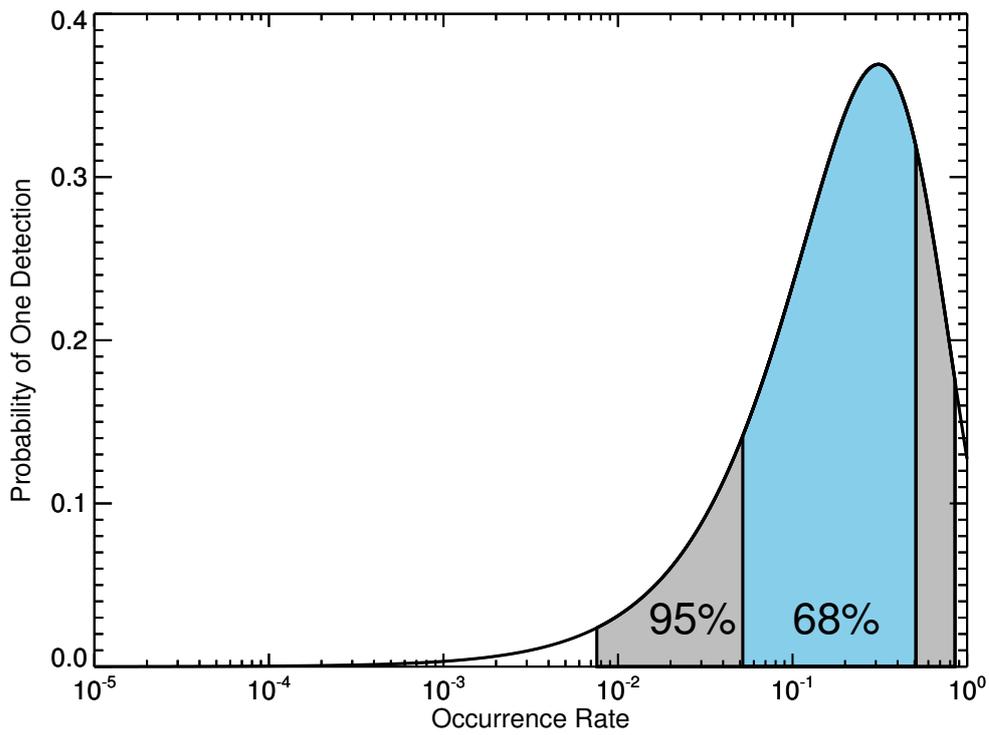

Figure S9: Probability distribution of occurrence rates of disintegrating planets around white dwarfs. The 68% and 95% confidence intervals are shaded in blue and grey, respectively. Because there is only one known object, the constraint on occurrence rate is weak.

28